\documentclass[letterpaper]{article} 
\usepackage{aaai2026}
\usepackage{times}  
\usepackage{helvet}  
\usepackage{courier}  
\usepackage[hyphens]{url}  
\usepackage{graphicx} 
\usepackage{natbib}  
\usepackage{caption} 
\usepackage{algorithm}
\usepackage{algorithmic}

\usepackage{amsmath}
\usepackage{amsfonts}
\usepackage{booktabs}
\usepackage{multirow}
\usepackage{array}
\usepackage{tabularx}

\usepackage{newfloat}
\usepackage{listings}
\DeclareCaptionStyle{ruled}{labelfont=normalfont,labelsep=colon,strut=off} 
\floatstyle{ruled}
\newfloat{listing}{tb}{lst}{}
\floatname{listing}{Listing}
\newif\ifsubmit

\usepackage{tabularx}
\usepackage{makecell}
\usepackage{multirow}
\usepackage{color}
\usepackage{graphicx}
\usepackage{booktabs}
\usepackage[table]{xcolor}
\usepackage{float}

\definecolor{skyblue}{RGB}{135,206,235}
\definecolor{aliceblue}{RGB}{240,248,255}
\definecolor{alicecolor}{RGB}{250,26,150}
\definecolor{bobcolor}{RGB}{0,128,0}
\definecolor{carolcolor}{RGB}{255,140,0}
\definecolor{davecolor}{RGB}{199,21,133}
\definecolor{evecolor}{RGB}{0,102,204}

\ifsubmit

\else

\fi

\title{Emotion-Qwen:~A~Unified Framework for Emotion and Vision Understanding}
\author{
    Dawei Huang\textsuperscript{\rm 1}, Qing Li\textsuperscript{\rm 1}, Chuan Yan\textsuperscript{\rm 2}, Zebang Cheng\textsuperscript{\rm 1}, 
    Zihao Han\textsuperscript{\rm 3}, Yurong Huang\textsuperscript{\rm 4}, Xiang Li\textsuperscript{\rm 1}, Bin Li\textsuperscript{\rm 5}, 
    Xiaohui Wang\textsuperscript{\rm 6}, Zheng Lian\textsuperscript{\rm 7}, Zhi-Qi Cheng\textsuperscript{\rm 3}, Xiaojiang Peng\textsuperscript{\rm 1}\thanks{Corresponding author.}
}
\affiliations{
    \textsuperscript{\rm 1}Shenzhen Technology University \quad
    \textsuperscript{\rm 2}Stanford University \quad
    \textsuperscript{\rm 3}University of Washington \\
    \textsuperscript{\rm 4}University of Electronic Science and Technology of China \quad
    \textsuperscript{\rm 5}Skyworth Digital \quad
    \textsuperscript{\rm 6}Shenzhen Xiaopai Tech Co. \\
    \textsuperscript{\rm 7}Institute of Automation, Chinese Academy of Sciences
}

\usepackage{bibentry}

\begin{document}

    \maketitle

\begin{figure*}[!h]
\centering
\includegraphics[width=\linewidth]{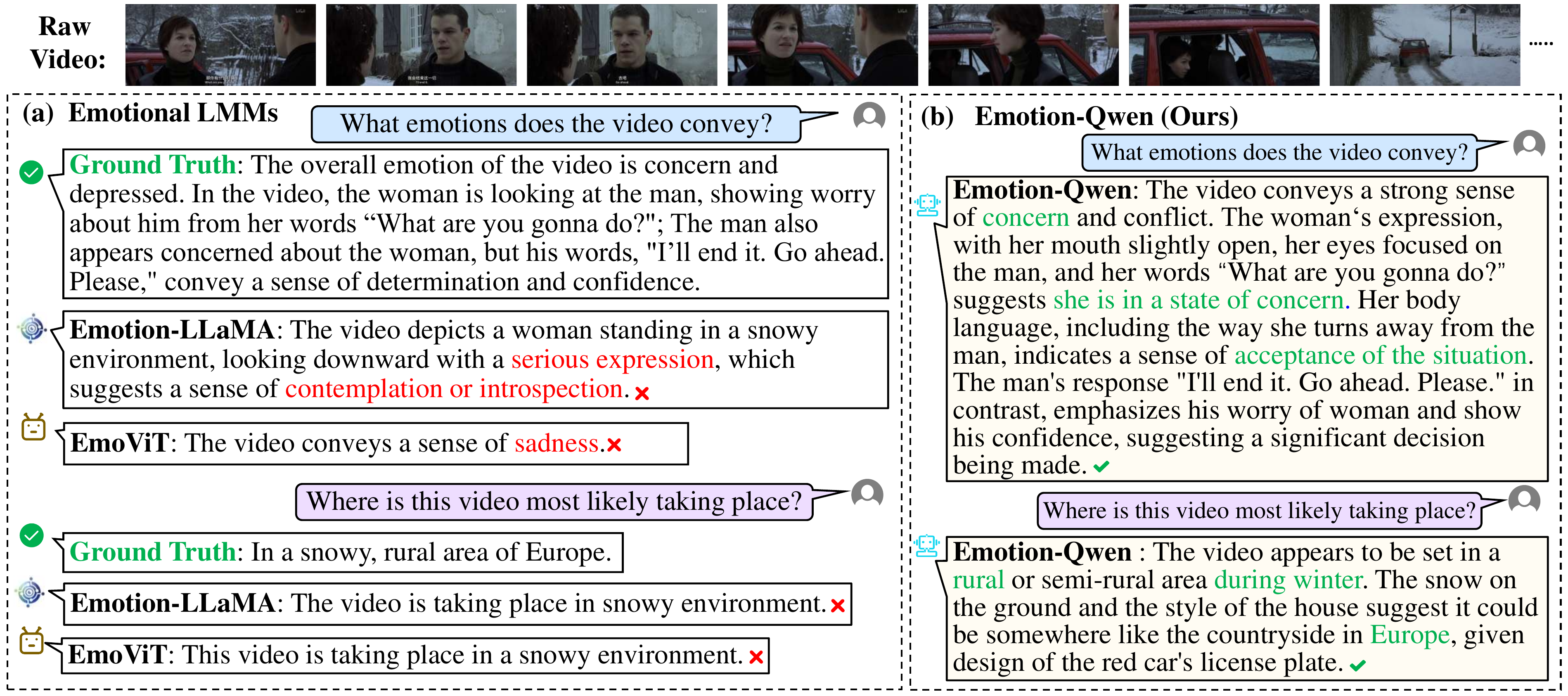}
\vspace{-0.2in}
\caption{\small \textbf{Motivation behind Emotion-Qwen (zoom in for detailed Q\&A):} (a) Current state-of-the-art emotional Large Multimodal Models (LMMs) experience severe catastrophic forgetting, causing inaccurate emotion recognition, ineffective dialogue comprehension, and limited multimodal reasoning abilities. (b) In contrast, Emotion-Qwen effectively addresses these limitations, balancing fine-grained emotion reasoning with robust general vision-language understanding. Incorrect outputs are marked in red; correct outputs in green.}
\label{fig:first fig}
\vspace{-0.2in}
\end{figure*}

\begin{abstract}
Accurate emotion understanding in videos necessitates effectively recognizing and interpreting emotional states by integrating visual, textual, auditory, and contextual cues. Although recent Large Multimodal Models (LMMs) have exhibited significant progress in general vision–language (VL) tasks, their performance often deteriorates in emotion-specific scenarios, exhibiting catastrophic forgetting when fine-tuned on emotion-centric tasks. To overcome these limitations, we propose \textbf{Emotion-Qwen}, a unified multimodal framework designed to simultaneously enable robust emotion understanding and preserve general VL reasoning capabilities. Emotion-Qwen introduces a novel Hybrid Compressor based on a Mixture-of-Experts (MoE) architecture, dynamically routing inputs to optimally balance emotion-specific processing and general multimodal reasoning. We further propose a carefully structured three-stage pre-training pipeline, leveraging extensive general and emotion-focused datasets to strengthen multimodal representation robustness and model adaptability. Additionally, we develop the \emph{Video Emotion Reasoning (VER)} dataset, a large-scale bilingual resource containing over 40K video clips annotated with detailed context-aware emotional descriptions, significantly facilitating research on fine-grained emotional reasoning. Extensive experiments confirm that Emotion-Qwen achieves state-of-the-art performance across multiple emotion recognition and reasoning benchmarks, while maintaining highly competitive results in general VL tasks.
\end{abstract}

\noindent\textbf{Code} -- \url{https://anonymous.4open.science/r/Emotion-Qwen-Anonymous}

\begin{figure*}
\centering
\includegraphics[width=\textwidth]{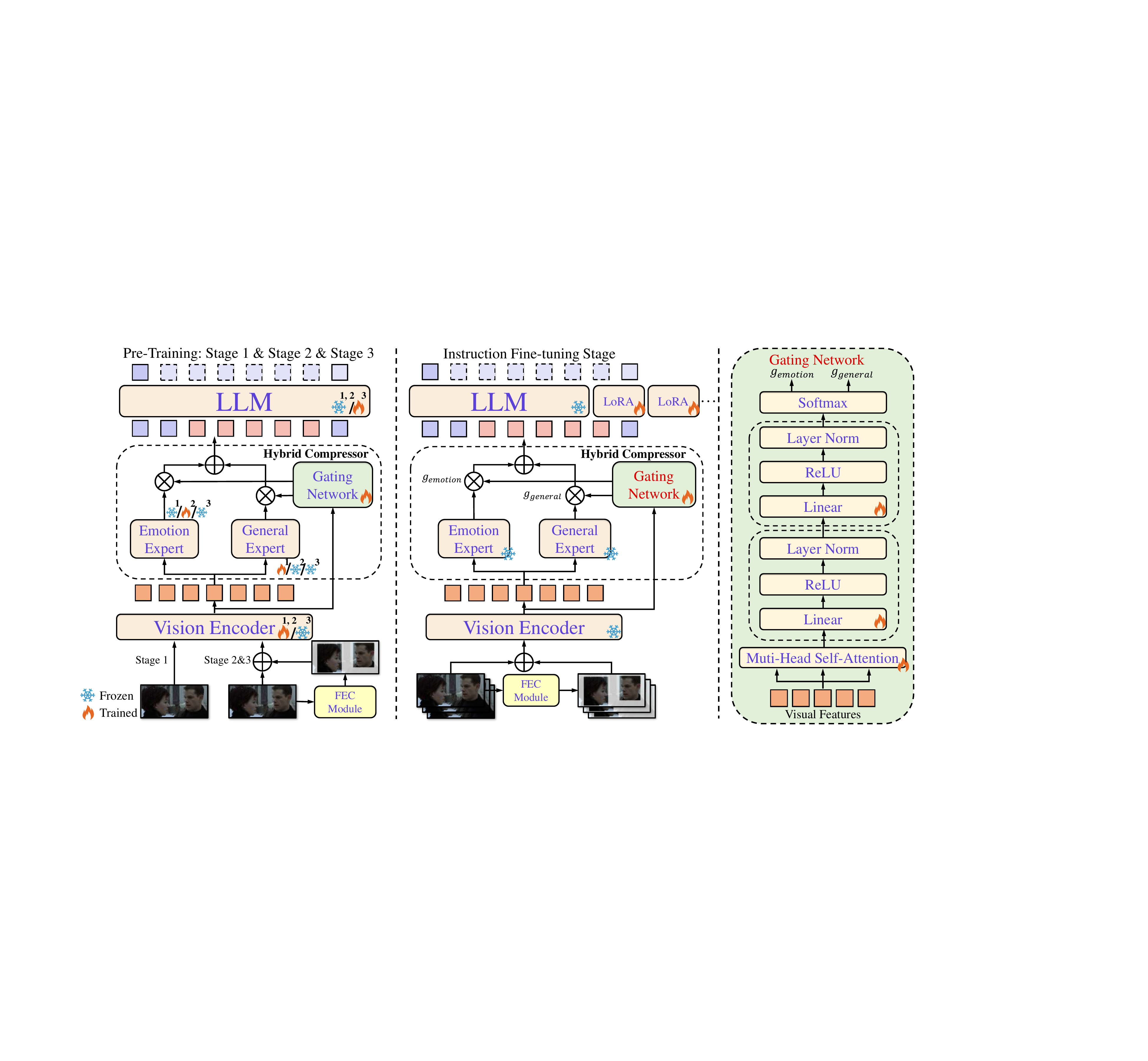}
\vspace{-0.15in}
\caption{\small Overview of Emotion-Qwen’s overall training pipeline and model architecture. The Facial Emotion Capture (FEC) Module identifies key emotional cues from input videos, while the Hybrid Compressor employs an attention-based Gating Network to dynamically integrate emotion-specific and general multimodal features. The detailed structure of the Gating Network is illustrated on the right.}
\vspace{-0.2in}
\label{fig:Training}
\end{figure*}

\section{Introduction}
Emotion understanding is a foundational yet challenging task in affective computing, necessitating seamless integration of multimodal cues—including facial expressions, vocal intonations, linguistic subtleties, and contextual visual information—to accurately interpret complex emotional states. Although early research in \textit{facial expression recognition}\cite{Attention-Driven-Xiaojiang, VGGFace2}, \textit{audio emotion recognition}\cite{wav2vec2}, and \textit{text sentiment analysis}~\cite{InstructERC} advanced unimodal emotion analysis, these approaches often struggle to capture the rich, dynamic, and contextually nuanced nature of human emotions.

Recent advancements in Large Multimodal Models (LMMs)\cite{FromLLMtoLMM, Qwen2-VL, MiniCPM-V} have shown promising potential for enhancing multimodal emotion understanding. Initially developed for general vision-language tasks—such as visual question answering\cite{ScienceQA, TextVQA}, cross-modal reasoning~\cite{MMBench, MME}, and image-text generation~\cite{POPE, OCR-VQA}—these models exhibit robust generalization across modalities. Nevertheless, existing LMMs continue to face significant challenges in specialized emotion reasoning scenarios~\cite{GPT-4vwithemotion}. Recent studies have addressed these limitations by fine-tuning LMMs specifically for multimodal emotion recognition (MER)~\cite{MER2024, DFEW}, leveraging integrated visual, auditory, and textual cues.

However, as illustrated in Figure~\ref{fig:first fig}, current methods have several critical shortcomings. Primarily, existing approaches are limited to coarse-grained classification of basic emotions, lacking deeper interpretability and context-aware emotional reasoning. Additionally, fine-tuning LMMs on emotion-centric datasets frequently leads to catastrophic forgetting~\cite{More_Than_Catastrophic_Forgetting, Catastrophic_Forgetting_in_PEFT}, significantly impairing their general vision-language performance. This trade-off between emotion specialization and multimodal generalization represents a substantial barrier to creating robust and emotionally intelligent AI systems.

To address these challenges, we introduce \textbf{Emotion-Qwen}, a unified multimodal framework explicitly designed to achieve balanced capabilities in both nuanced emotion understanding and general vision-language reasoning. Emotion-Qwen integrates a dedicated \textit{Facial Emotion Capture (FEC)} module with an attention-aware Mixture-of-Experts (MoE) \textit{Hybrid Compressor}. The FEC module effectively extracts expressive facial features, enhancing emotion-specific representation learning, while the Hybrid Compressor dynamically routes multimodal inputs between emotion-specialized and general-purpose experts. This design ensures comprehensive emotion modeling and efficient cross-task knowledge sharing, facilitating robust affective reasoning alongside general vision-language alignment.

To systematically evaluate and further enhance the sophisticated emotional reasoning capabilities of Emotion-Qwen, we propose the \textit{Video Emotional Reasoning (VER)} dataset—a large-scale, bilingual (Chinese-English) resource comprising over 40K video clips enriched with more than 80K detailed annotations. VER explicitly emphasizes contextual and causal emotional cues beyond traditional emotion classification, supporting advanced fine-grained emotional reasoning tasks. Extensive experiments demonstrate that Emotion-Qwen achieves state-of-the-art results across multiple emotion-centric and general vision-language benchmarks, consistently exhibiting superior reasoning abilities and scalable multimodal understanding.

Our contributions are summarized as follows:
\begin{itemize}
\item We propose Emotion-Qwen, the first multimodal framework explicitly designed to simultaneously balance specialized emotion understanding and general vision-language reasoning, significantly reducing catastrophic forgetting in emotion-adapted LMMs.
\item Emotion-Qwen uniquely integrates a Facial Emotion Capture (FEC) module with a Hybrid Compressor, enabling efficient extraction of expressive emotional features and adaptive multimodal representation alignment.
\item Leveraging our carefully designed training strategy and curated datasets, Emotion-Qwen achieves state-of-the-art performance on several key benchmarks, including 87.3 on MMBench, 87.9 on TextVQA (zero-shot), and—following instruction tuning—78.31 UAR on DFEW, 8.25/8.16 Clue/Label Overlap scores on EMER, and 85.49 accuracy on EmoSet.
\end{itemize}

\section{Related Work}

\noindent \textbf{Large Multimodal Models (LMMs)}\cite{LLaVA, Qwen-VL, Qwen2-VL, MiniCPM-V, GPT-4} have become a leading approach for integrating large language models (LLMs)\cite{LLaMA, LLaMA2, Qwen} with multiple modalities, enabling unified reasoning across visual, textual, and auditory inputs. Recent advancements have substantially improved general vision-language (VL) tasks, such as visual question answering, image captioning, and cross-modal retrieval~\cite{MMBench, VQAv2, TextVQA}. However, existing LMMs still exhibit significant limitations in fine-grained emotion reasoning. Evaluations of prominent models such as GPT-4V highlight persistent difficulties in interpreting nuanced emotions and providing causal explanations within complex video scenarios~\cite{GPT-4vwithemotion}. Although recent efforts attempt to fine-tune LMMs specifically on emotion datasets~\cite{Emotion-LLaMA, EmoVIT}, these approaches frequently lead to catastrophic forgetting, thereby compromising general VL capabilities.

\noindent \textbf{Multimodal Emotion Recognition (MER)} aims to accurately identify emotional states by integrating visual, auditory, and textual information. Traditional MER methods primarily aggregate unimodal features for discrete emotion classification tasks, exemplified by datasets like DFEW~\cite{DFEW} and MELD~\cite{MELD}. While emphasizing facial expressions and conversational contexts, these datasets typically provide limited, single-label annotations, lacking detailed contextual information and deeper reasoning insights. Recent advancements introduce novel architectures and training strategies to enhance model robustness and generalization. Benchmarks such as MER2023 and MER2024~\cite{MER2023, MER2024} emphasize semi-supervised learning, open-vocabulary emotion classification, and robustness to noisy data. Models like Emotion-LLaMA~\cite{Emotion-LLaMA} and AffectGPT~\cite{AffectGPT-new} employ instruction tuning and multi-stage training to significantly improve emotion reasoning, though often suffer from task-specific overfitting. Additionally, newer datasets like EMER~\cite{EMER} provide richer explanatory annotations but remain limited to isolated contexts. To overcome these limitations, we propose the Video Emotional Reasoning (VER) dataset, explicitly capturing dynamic emotional expressions, narrative contexts, and cross-situational inference, thereby significantly enhancing multimodal emotion reasoning capabilities.

\section{Methodology}

\subsection{Overview of Emotion-Qwen}
As illustrated in Figure~\ref{fig:Training}, Emotion-Qwen consists of four key components: the Facial Emotion Capture (FEC) Module, Vision Encoder, Hybrid Compressor, and the LLM Backbone. The FEC Module identifies and extracts critical frames showcasing prominent emotional expressions from video inputs. Subsequently, we utilize a pre-trained CLIP ViT~\cite{ViT, Qwen2-VL} as the vision encoder, transforming the visual pixel values $\mathbf{P} \in \mathbb{R}^{H \times W \times 3}$ into rich visual embeddings $\mathbf{E} = f_{\text{ViT}}(\mathbf{P})$.

Next, the Hybrid Compressor efficiently compresses and aligns these visual embeddings $\mathbf{E} \in \mathbb{R}^{N_1 \times d_v}$ into compact visual tokens $\mathbf{V} \in \mathbb{R}^{N_2 \times d_t}$, synchronizing them with the textual feature representations $\mathbf{T} \in \mathbb{R}^{M \times d_t}$. Here, $d_v$ and $d_t$ denote the dimensions of the visual and textual features, respectively; $N_1$ represents the length of the visual embeddings, $N_2$ indicates the number of compressed visual tokens, and $M$ corresponds to the length of text tokens. Finally, these integrated visual tokens $\mathbf{V}$ and textual tokens $\mathbf{T}$ are fed into the Qwen2.5~\cite{Qwen2.5} large language model backbone, generating the multimodal output $\hat{\mathbf{Y}} = f_{\text{LLM}}(\mathbf{V}, \mathbf{T}, \mathbf{Prompt})$.

\subsection{Facial Emotion Capture Module}
To effectively extract emotion-relevant facial expressions from video inputs, we propose the Facial Emotion Capture (FEC) Module, built upon the open-source DeepFace framework~\cite{DeepFace}. Given the $i$-th video frame $\mathbf{F}_i \in \mathbb{R}^{h \times w \times 3}$, with $h$ and $w$ denoting its height and width respectively, we initially perform face detection to acquire bounding boxes $\mathbf{B}_j$.
For each identified facial region $\mathbf{F}_{ij}$ in frame $i$, DeepFace computes facial landmarks $\mathbf{A}_{ij} = g(\mathbf{F}_{ij})$ and predicts emotion probability vectors $\mathbf{E}_{ij} = h(\mathbf{F}_{ij})$, where $g(\cdot)$ and $h(\cdot)$ represent landmark localization and emotion classification functions, respectively.
Subsequently, frames exhibiting high-confidence predictions across emotion categories are selected as key emotional frames. To emphasize expressive facial cues and reduce extraneous background noise, spatial masking is applied, preserving only salient facial regions.
Finally, these key frames are temporally ordered and concatenated with the original video stream, generating an enriched sequence that accentuates emotionally relevant content while retaining the temporal coherence of the input. This enhanced sequence is then forwarded to the vision encoder for subsequent multimodal processing.

\subsection{Hybrid Compressor with Expert Routing}
\label{subsec:Hybrid Compressor}
Visual embeddings often contain redundant or irrelevant information, increasing computational overhead and potentially introducing feature noise. To address these challenges, we propose a Hybrid Compressor (HC) that dynamically compresses and aligns visual embeddings by adaptively routing information through specialized experts.

Specifically, HC employs two expert modules—an \textit{Emotion Expert} and a \textit{General Expert}—alongside an attention-based \textit{Gating Network} that dynamically weights and combines expert outputs. Both experts are implemented via multi-layer perceptrons (MLPs) integrated with GELU activation~\cite{GELU} and layer normalization. Given input visual embeddings $\mathbf{E} \in \mathbb{R}^{N \times d_v}$, the Emotion Expert computes representations as:
\begin{equation}
\mathbf{V}_{\text{emo}} = \sigma\left(\mathbf{W}_{\text{emo}} \cdot \text{GELU}(\mathbf{W}'_{\text{emo}} \cdot \mathbf{E} + \mathbf{b}'_{\text{emo}}) + \mathbf{b}_{\text{emo}}\right),
\label{eq:emo_expert_representation}
\end{equation}
while the General Expert computes:
\begin{equation}
\mathbf{V}_{\text{gen}} = \sigma\left(\mathbf{W}_{\text{gen}} \cdot \text{GELU}(\mathbf{W}'_{\text{gen}} \cdot \mathbf{E} + \mathbf{b}'_{\text{gen}}) + \mathbf{b}_{\text{gen}}\right),
\label{eq:gen_expert_representation}
\end{equation}
where $\sigma(\cdot)$ denotes the layer normalization function, and parameters $\mathbf{W}, \mathbf{b}$ are learned during training.

To adaptively fuse the experts, the Gating Network leverages an attention-based mechanism to calculate input-dependent weights:
\begin{equation}
\mathbf{G} = \text{softmax}\left(\mathbf{W}_{\text{gate}} \cdot \text{Attention}(\mathbf{E}) + \mathbf{b}_{\text{gate}}\right).
\label{eq:gate}
\end{equation}
The final output visual embedding $\mathbf{V}_{\text{out}}$ is obtained via element-wise gating of both experts:
\begin{equation}
\mathbf{V}_{\text{out}} = \mathbf{G} \odot \mathbf{V}_{\text{emo}} + (\mathbf{1} - \mathbf{G}) \odot \mathbf{V}_{\text{gen}}.
\label{eq:MoE}
\end{equation}
This dynamic routing strategy effectively preserves task-relevant information, reduces redundancy, and enhances multimodal representation alignment, significantly benefiting emotional reasoning and general vision-language tasks.

\begin{table}[!t]
    \small
  \begin{tabularx}{\columnwidth}{c|>{\centering\arraybackslash}X |c}
    \toprule
     Stage&Datasets&Size\\
    \midrule
    \multirow{2}{*}{1} & ImageNet, SBU, COCO-Caption, CC12M, VQAv2 & \multirow{2}{*}{15.1M}\\
    \midrule
    \multirow{1}{*}{2} & LAION-Face-20M, RAF-DB & \multirow{1}{*}{10.2M}\\ 
    \midrule
    \multirow{2}{*}{3} & LLaVA-mix665K, LLaVAR, OCR-VQA, GQA, OKVQA & \multirow{2}{*}{1.9M}\\
  \bottomrule
    \end{tabularx}
    \vspace{-0.1in}
  \caption{\small Overview of publicly available pre-training datasets. Stage 1 uses image-text pairs to align vision and language modalities. Stage 2 introduces emotion-centric data spanning diverse cultural contexts. Stage 3 incorporates OCR and instruction-tuned datasets to strengthen reasoning in complex vision-language tasks.}
    \label{tab:Pre-training data}
    \vspace{-0.2in}
\end{table}

\subsection{Video Emotion Reasoning Dataset}
\label{sec:Video Emotion Reasoning Dataset}
The \textit{VER} dataset was constructed leveraging publicly available data from MAFW \cite{MAFW} and MER2024 \cite{MER2024}, selecting 8,034 and 36,357 annotated samples, respectively.
For MER2024, we employed a two-stage filtering approach following previous work \cite{AffectGPT-new}: first, we applied TalkNet \cite{TalkNet} to identify and remove mismatches between audio and visual components; subsequently, we conducted a model-guided, human-assisted refinement. For MAFW, we specifically retained samples containing clear emotion annotations accompanied by explicit facial action descriptions, providing reliable supervision signals.

\begin{figure}[!t]
  \centering
  \includegraphics[width=1.0\linewidth]{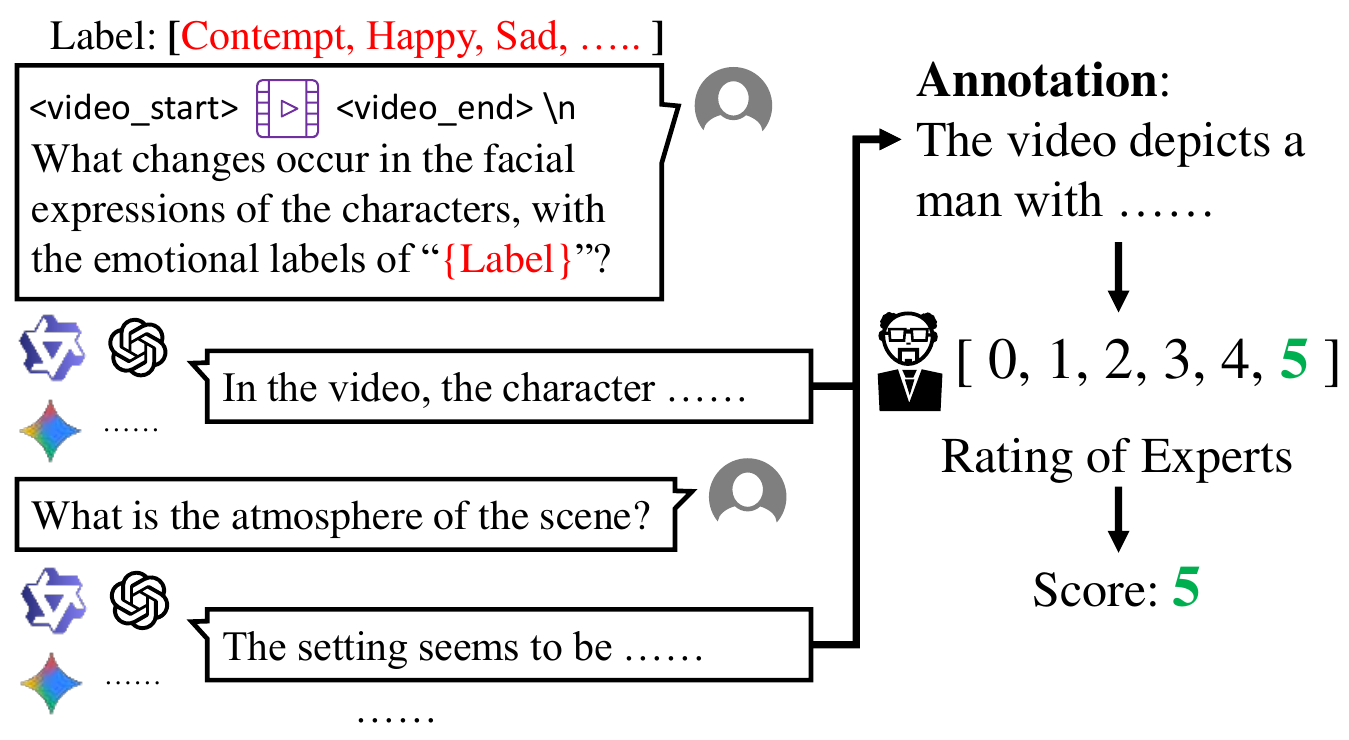}
  \vspace{-0.2in}
  \caption{\small Construction pipeline of the Video Emotion Reasoning (VER) dataset. Outputs from multiple models are synthesized and human-verified to reduce bias and hallucination.}
  \label{fig:construction of VER}
\vspace{-0.2in}
\end{figure}

As illustrated in Figure~\ref{fig:construction of VER}, to minimize potential biases and illusions from large models, we utilized robust closed-source LMMs, including Qwen-VL-Max \cite{Qwen2-VL} and GPT-4 \cite{GPT-4}. These models were provided with carefully crafted prompts along with coarse emotion labels, guiding them to pinpoint key emotional cues such as facial expressions and contextual settings.
To further mitigate model-induced biases, outputs from these models were integrated and synthesized using DeepSeek \cite{Deepseekv3}, producing fine-grained emotional annotations. Subsequently, 12 professional analysts rigorously evaluated the annotations on a 0-to-5 scale. Only annotations rated 3 or higher were incorporated into the final \textit{VER} dataset.
Unlike previous datasets, \textit{VER} uniquely emphasizes detailed emotional reasoning, explicitly capturing facial expressions, body language, and rich visual context. Additionally, responses from multiple models were cross-validated, and emotion-domain experts independently assessed the annotations’ relevance and accuracy.

\begin{table*}[!h]
    \small
    \centering
    \setlength{\tabcolsep}{1.0pt}
    \begin{tabular}{lcc|ccccccc|ccccccc}
    \toprule
    \multirow{4}{*}{Model} & \multirow{4}{*}{Modal} & \multirow{4}{*}{Size} & \multicolumn{7}{c}{General Benchmarks} & \multicolumn{7}{|c}{Emotion Tasks} \\
    \cmidrule{4-17}
     & & & \multirow{2}{*}{MME} &  MM- &  \multirow{2}{*}{POPE} &  Science- &  Seed- &  Text- &  \multirow{2}{*}{VQAv2} & \multicolumn{2}{c}{MER2024} & \multicolumn{2}{c}{DFEW} &\multicolumn{2}{c}{EMER} & \multirow{2}{*}{EmoSet}\\
    & & &  & Bench &  & QA & Bench & VQA &  & SEMI & NOISE & WAR & UAR & CLUE & LABEL \\
    \midrule
    \multicolumn{14}{l}{\textbf{Emotion LMMs}} \\
        EmoViT  & V&7B & 741.3& 38.2& 66.7& -&37.3 &15.9 &25.8 &34.78 & 31.35&34.42 & 17.21&2.67&3.62& \textbf{83.36} \\
    Emotion-LLaMA   & A,V,T & 7B & 1538.5 & 81.2 & 81.3 & 42.3 & 40.6 & 26.7 & 71.4 & 73.62 & 73.62 & 77.06 & \textbf{64.21} & 7.83 & 6.25 & 43.01  \\
    \midrule
    \textbf{General LMMs} &\\
    GPT-4V & V,T & - & 2070.2 & 75.0 & 81.8 & - &  71.6 & 78.0 & - & - & - & 55.00 & 36.96 & - &-&-  \\
    LLaVA-v1.5  & V,T & 7B & 1823.3 & 63.3 & 86.1 & 65.2 & 59.3 &  51.0 & 84.3 & 34.97& 31.74 & 35.33& 17.66 & 3.36& 4.67&59.00  \\
    InstructBLIP & V,T & 13B & 1504.6& -& -& 63.1& 58.8& 50.7 & 65.0 & 22.10 & 19.18&26.61& 13.31&2.55 & 4.12&58.79\\
    Qwen-VL-Chat  & V,T & 7B & 1848.3 & 61.8 & 79.9 &67.1 & 65.4 & 61.5 & 78.2  & 39.95 & 35.76 & 30.97 & 15.48& 3.71 & 4.89 &50.85\\
    Qwen2-VL  & V,T & 7B & \textbf{2326.8} & 83.0 & 86.2 & \textbf{78.1} & \textbf{81.8} & 84.3 & \textbf{84.9} & 56.18 & 56.08 & 63.86 & 31.93 & 4.34 & 5.59 &55.89  \\
    DeepSeek-VL  & V,T & 7B & 1765.4 & 73.2 & \textbf{88.1} & 57.3 & 70.4 & 64.7 & 52.9 & 20.52& 20.12 &37.51& 18.75 & 3.64 & 5.08 & 44.53 \\
    \midrule
    \rowcolor[RGB]{222, 222, 222}
     Emotion-Qwen(pretrained)  & V,T& 7B & 2163.5 & \textbf{87.3} & 83.2 & 77.2 & 71.5  & \textbf{87.9} & 84.8 & \textbf{82.85} & \textbf{77.53} & \textbf{77.19} & 38.60 & 6.49 & \textbf{6.81} & 81.54 \\
     \rowcolor[RGB]{230, 242, 255}
     Emotion-Qwen(fine-tuned) & V,T& 7B & 2054.8 &  85.3& 79.4  & \textbf{80.7} &68.4& 86.3 & 79.2 &\textbf{85.47} & \textbf{79.67} & \textbf{78.31} & 62.11 & \textbf{8.25} & \textbf{8.16} & \textbf{85.49}\\
    \bottomrule
    \end{tabular}
    \vspace{-0.1in}
    \caption{\small Comparison of Emotion-Qwen with leading models on general benchmarks and emotion-related tasks. The second column denotes input modalities: “V” for video, “T” for text, and “A” for audio. Dataset-specific abbreviations indicate evaluation settings: “SEMI” and “NOISE” refer to MER2024 tracks; “WAR” and “UAR” are weighted and unweighted average recall; “CLUE” and “LABEL” represent Clue and Label Overlap scores in EMER~\cite{EMER}, ranging from 0 to 10 and assessed by ChatGPT. Best results are shown in bold.}
    \label{tab:zero-shot comparison}
    \vspace{-0.1in}
\end{table*}

\subsection{Pre-training for Emotion-Qwen}
\label{sec:Pre-training}
During pre-training, we follow established approaches~\cite{Qwen-VL, MiniCPM-V}, aggregating publicly available datasets as detailed in Table~\ref{tab:Pre-training data}. As illustrated in the left portion of Figure~\ref{fig:Training}, this process consists of three distinct stages:

\noindent \textbf{Stage 1: Warm-up of General Expert}.
We randomly initialize the Hybrid Compressor and train the General Expert, Gating Network, and Vision Encoder using a general image-text corpus, keeping the LLM backbone frozen and the Facial Emotion Capture Module inactive. This stage aligns the General Expert with the vision encoder and LLM, preparing it to effectively handle generic visual features.

\noindent \textbf{Stage 2: Warm-up of Emotion Expert}.
In this stage, we train the Emotion Expert, Gating Network, and Vision Encoder on facial emotion datasets encompassing diverse cultural expressions. The primary goal is to enable the Emotion Expert to learn effective and robust emotional representations.
Collectively, Stages 1 and 2 warm up the Hybrid Compressor, ensuring that both expert modules function optimally within their respective domains.

\noindent \textbf{Stage 3: Fine-tuning on Vision-Language Tasks}.
Finally, we unlock the LLM parameters and fine-tune the model using diverse instruction-based datasets. This stage enables Emotion-Qwen to learn complex multimodal patterns, facilitating comprehensive generalization and sophisticated vision-language understanding.

\subsection{Emotional Instruction Fine-tuning}
\label{sec:Instruction Fine-tuning}
In this stage, we utilize our constructed \textit{VER} dataset along with additional emotion-focused datasets, such as DFEW~\cite{DFEW} and EmoViT~\cite{EmoVIT}, for instruction-based fine-tuning. Specifically, we freeze the Vision Encoder and Hybrid Compressor modules, applying multiple instances of Low-Rank Adaptation (LoRA) to selectively fine-tune the LLM backbone with task-specific emotional prompts. This multi-LoRA approach allows Emotion-Qwen to capture nuanced dataset-specific emotional characteristics, significantly enhancing its performance across diverse multimodal emotion reasoning tasks.

\section{Experiments}
\label{sec:Experiment}
We evaluate Emotion-Qwen across a range of general vision-language and emotion-related tasks to comprehensively assess its performance. Details of the datasets and benchmarks are provided in the \textit{Supplementary Material}.

\begin{table}[h]
    \small
    \centering
    \setlength{\tabcolsep}{1pt} 
    \begin{tabular}{ l | c c c | c }
    \toprule
    \multirow{2}{*}{H-Params} & \multicolumn{3}{c|}{Pre-training} & \multirow{2}{*}{Fine-tuning}  \\
    \cmidrule{2-4} 
     & Stage 1 & Stage 2 & Stage 3   \\
    \midrule
    Optimizer       & \multicolumn{3}{c|}{AdamW}   & AdamW     \\
    Learning Rate   & \multicolumn{3}{c|}{$1 \times 10^{-6}$} & $1 \times 10^{-4}$        \\
    LR Schedule  & \multicolumn{3}{c|}{Cosine}      & Cosine     \\
    Weight Decay    & \multicolumn{3}{c|}{0.1}     & 0.1       \\
    Adam $\beta_2$         & \multicolumn{3}{c|}{0.95}    & 0.95      \\
    Warm-up Ratio   & \multicolumn{3}{c|}{0.01}    & 0.01      \\
    Epochs  & 1      & 1       & 3      & 5         \\
    Max Image Resolution    & \multicolumn{3}{c|}{$1280 \times 784$} & $1280 \times 784$         \\
    Max Video Resolution   & \multicolumn{3}{c|}{-}  & $448 \times 448$   \\
    LoRA Rank ($r$)        & \multicolumn{3}{c|}{-}      & 64       \\
    LoRA Scaling ($\alpha$)& \multicolumn{3}{c|}{-}      & 64       \\
    LoRA Dropout    & \multicolumn{3}{c|}{-}      & 0.05     \\
    DeepSpeed  & \multicolumn{3}{c|}{Zero2}      & Zero2     \\
    \bottomrule
    \end{tabular}
    \vspace{-0.1in}
    \caption{\small Summary of Emotion-Qwen's hyperparameter settings used in pre-training and instruction fine-tuning stages.}
    \label{tab:Hyperparameter}
    \vspace{-0.1in}
\end{table}

\subsection{Implementation Details}
\label{subsec:Implementation_Details}
Following the pre-training pipeline, Emotion-Qwen was trained using 3 NVIDIA A800 80GB GPUs with DeepSpeed~\cite{deepspeed} for distributed training. To reduce memory and computation overhead, we incorporated FlashAttention2~\cite{FlashAttention-2} and constrained the maximum image resolution to $1280 \times784$. Consistent with prior work~\cite{Qwen2-VL, MiniCPM-V}, cross-entropy loss was used as the training objective.
In the instruction fine-tuning stage, we employed multiple LoRA adapters to fine-tune the LLM backbone for emotion understanding. Input videos were resized to $448 \times448$ and sampled at 3 FPS to balance temporal coherence and memory usage. Each LoRA was trained independently on its corresponding dataset for 5 epochs to prevent overfitting. A complete summary of hyperparameters is provided in Table~\ref{tab:Hyperparameter}.

\subsection{Zero-shot Evaluation}
We evaluated Emotion-Qwen’s zero-shot capabilities across several widely-used benchmarks. As summarized in Table~\ref{tab:zero-shot comparison}, Emotion-Qwen (the gray-highlighted row) consistently outperforms most open-source LMMs and achieves competitive performance even compared to proprietary models such as GPT-4V. Specifically, Emotion-Qwen surpasses leading models like GPT-4V, Qwen2-VL, and DeepSeek-VL on general benchmarks, achieving scores of 87.3 on MMBench, 77.2 on ScienceQA, and 87.9 on TextVQA.

Notably, Emotion-Qwen demonstrates robust performance in challenging multimodal emotion recognition tasks, achieving a UAR of 77.19 on the DFEW dataset without fine-tuning, closely approaching current state-of-the-art (SOTA) results (Table~\ref{tab:lora-finetuning-results}). Moreover, Emotion-Qwen achieves a Clue Overlap score of 6.49 on the EMER emotion reasoning task, even without audio input, surpassing general LMMs. Additionally, it sets a new SOTA for the Label Overlap metric with a score of 6.81, outperforming specialized emotion-focused LMMs such as Emotion-LLaMA.

\begin{figure}[t]
  \centering
  \includegraphics[width=\linewidth]{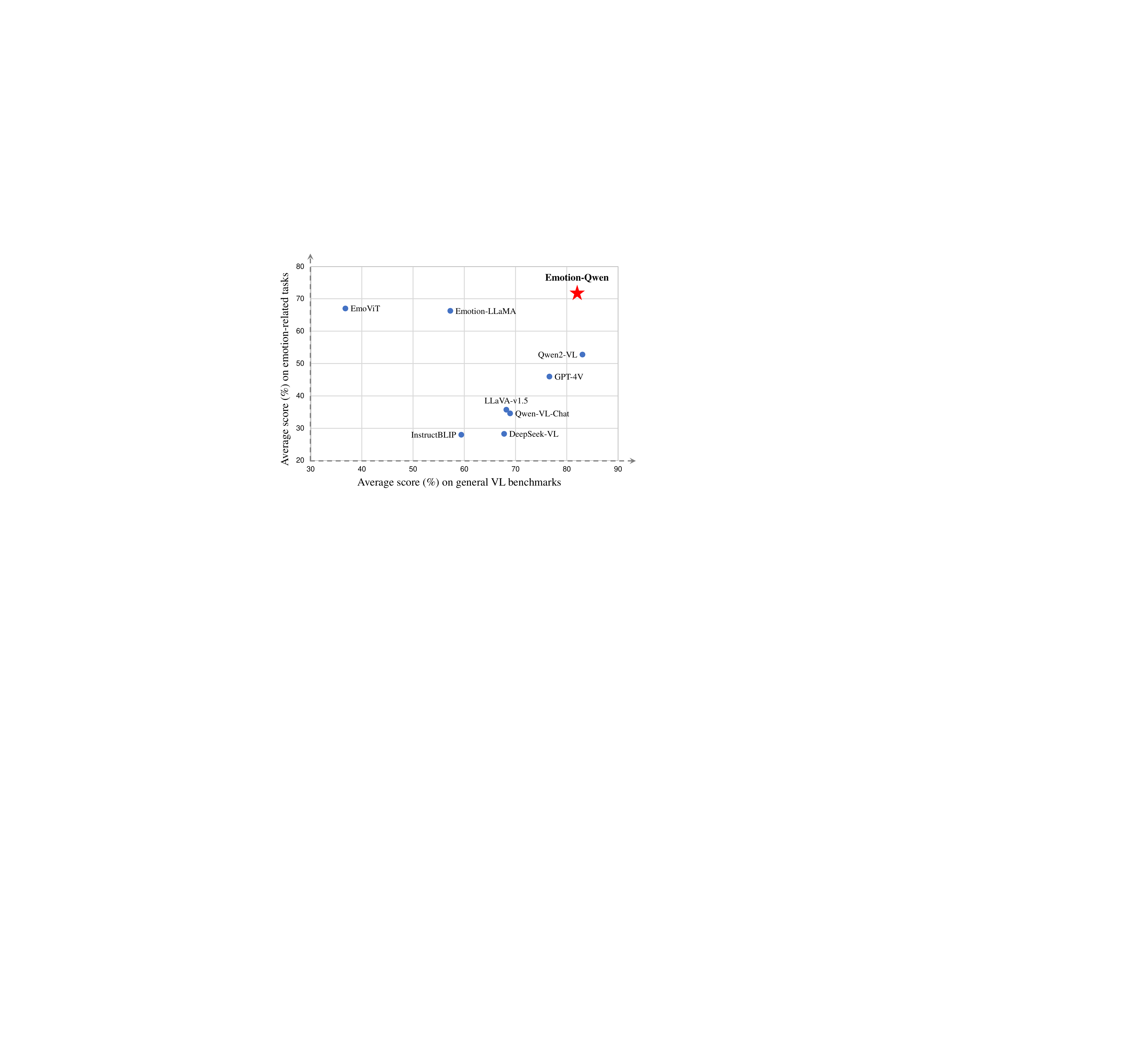}
  \vspace{-0.3in}
  \caption{\small Balanced performance of Large Multimodal Models (LMMs). Our Emotion-Qwen demonstrates a superior balance between fine-grained emotion reasoning and general vision-language understanding. Evaluation details are provided in Table~\ref{tab:zero-shot comparison}.}
  \label{fig:Scatter plot}
\end{figure}

\begin{table}[t]
\small
    \setlength{\tabcolsep}{1.0pt} 
    \begin{tabular}{l|c|c|c|c}
    \toprule
    \multirow{2}{*}{Model}  & \multicolumn{1}{c|}{MER2024} & \multicolumn{1}{c|}{DFEW} & \multicolumn{1}{c|}{EMER} & \multirow{2}{*}{EmoSet} \\
     & SEMI/NOISE  & WAR/UAR & CLU/LAB& \\
    \midrule
    MMA-DFER  & -/- &77.51*/\textbf{67.01}* &-/- &-\\
    AffectGPT & 78.80/78.80  & -/-  & -/- &-\\
    EmoViT  & 34.78/31.35 & 34.42/17.21 & 2.67/3.62 &83.36*\\
    Emotion-LLaMA & 73.62/73.62 & 77.06/64.21 & 7.83*/6.25 &43.01\\
    \midrule
    Emotion-Qwen& \textbf{85.47}/\textbf{79.67} & \textbf{78.31}/62.11 & \textbf{8.25}/\textbf{8.16} & \textbf{85.49} \\
    \bottomrule
    \end{tabular}
    \vspace{-0.1in}
    \caption{\small Performance comparison between fine-tuned Emotion-Qwen and existing state-of-the-art (SOTA) models on multimodal emotion benchmarks. $^*$ denotes previously reported SOTA results.}
    \label{tab:lora-finetuning-results} 
    \vspace{-0.1in}
\end{table}

\subsection{Instruction Fine-tuning Evaluation}
\label{Instruction_Fine_tuning_Evaluation}
To comprehensively evaluate Emotion-Qwen’s performance on emotion-centric tasks, we conducted instruction fine-tuning using our \textit{VER} dataset along with other specialized emotion datasets~\cite{DFEW, EmoVIT}. As shown in Table~\ref{tab:zero-shot comparison}, Emotion-Qwen (the blue-highlighted row)  achieves substantial improvements on emotion-related benchmarks post fine-tuning, while maintaining robust general vision-language capabilities, thereby minimizing catastrophic forgetting.

Furthermore, Table~\ref{tab:lora-finetuning-results} explicitly compares Emotion-Qwen against cutting-edge emotion-specific models, demonstrating its superiority across various tasks. Specifically, Emotion-Qwen attains scores of 85.47 and 79.67 on the SEMI and NOISE tracks of MER2024, respectively, outperforming prior leading models. On the DFEW dataset, Emotion-Qwen establishes a new state-of-the-art WAR score of 78.31, exceeding the previous best result achieved by MMA-DFER (77.51). Similarly, for the EmoSet dataset, Emotion-Qwen achieves the highest recorded accuracy of 85.49, surpassing EmoViT’s prior SOTA. Moreover, on the EMER emotion reasoning benchmark, Emotion-Qwen sets new records with Clue and Label Overlap scores of 8.25 and 8.16, respectively, significantly outperforming the previous SOTA Emotion-LLaMA. These results clearly highlight the effectiveness of our \textit{VER} dataset in enhancing fine-grained emotion understanding and demonstrate Emotion-Qwen’s exceptional ability to interpret complex emotional contexts, firmly establishing it as the new state-of-the-art in multimodal emotion recognition and reasoning.

\subsection{Ablation Study}
\label{Ablation Study}
\textbf{Facial Emotion Capture Module.} We evaluate the effectiveness of the Facial Emotion Capture (FEC) module by comparing model performance across multiple benchmarks with (“w/ FEC”) and without (“w/o FEC”) this component. As shown in Table~\ref{tab:Ablation on FEC Module and Hybrid Compressor}, incorporating the FEC module substantially improves results on emotion-related tasks. In contrast, a slight performance gain is observed on MMBench when the FEC module is excluded, likely because general vision-language benchmarks emphasize scene-level understanding over facial expression analysis. While the module may introduce minor noise in such tasks, its overall impact remains negligible.
On emotion-focused benchmarks such as MER2024 and DFEW, the FEC-equipped model consistently outperforms the baseline. Specifically, we observe a 2.29\% improvement on EmoSet, a 1.92\% gain on DFEW, and a 2.36\% boost on MER2024. These results highlight the FEC module’s effectiveness in extracting expressive facial cues and demonstrate its utility in fine-grained emotion recognition within multimodal contexts.

\begin{table}[t]
    \small
    \setlength{\tabcolsep}{1.0pt}
    \begin{tabular}{l|ccc | c c}
        \toprule
         & \multicolumn{3}{c}{Emotion Tasks} & \multicolumn{2}{|c}{General Benchmarks} \\
        \cmidrule{2-6}
        \multirow{2}{*}{Method} & \multicolumn{1}{c}{MER2024} & \multicolumn{1}{c}{DFEW} & \multirow{2}{*}{EmoSet} & \multicolumn{1}{c}{MM-}   &  Text-\\
        & SEMI/NOISE & WAR/UAR& & \multicolumn{1}{c}{Bench} &VQA\\
        \midrule
        w/o FEC &  81.90/76.12 & 77.09/36.78 & 79.25 & \textbf{87.48} & 87.93 \\
        w/ FEC &  \textbf{82.85}/\textbf{77.53} & \textbf{77.19}/\textbf{38.60} & \textbf{81.54} & 87.34 & \textbf{87.98} \\
        \midrule
        MLP &80.72/76.32& 76.97/38.48 & 80.36 & 84.22 & 79.54 \\
        Fusion &  82.63/76.88 & \textbf{77.78}/\textbf{38.89} & 81.25 & 84.38 & 79.26 \\
        HC & \textbf{82.85}/\textbf{77.53} & 77.19/38.60 & \textbf{81.54}& \textbf{87.34} & \textbf{87.98} \\
        \bottomrule
    \end{tabular}
    \vspace{-0.1in}
    \caption{\small Ablation study analyzing the effectiveness of the Facial Emotion Capture (FEC) Module and Hybrid Compressor (HC).}
    \label{tab:Ablation on FEC Module and Hybrid Compressor}
    \vspace{-0.1in}
\end{table}

\begin{table}[t]
\small
    \centering
    \setlength{\tabcolsep}{2.0pt}
    \begin{tabular}{l|c|cc|cc}
    \toprule
    \multirow{2}{*}{Proj.} & \multirow{2}{*}{Params}  & \multicolumn{2}{c|}{MMBench}  & \multicolumn{2}{c}{MER2024} \\
     & &  Latency$\downarrow$ & Score$\uparrow$ &  Latency$\downarrow$ & Score$\uparrow$ \\
    \midrule
    MLP & 44.57M & 42.9 ms & 84.22 & 390.3 ms & 76.32  \\
    Fusion & 89.14M  & 43.7 ms & 84.38  & 398.9 ms & 76.88  \\
    HC & 95.74M & 43.2 ms & \textbf{87.34} & 393.8 ms & \textbf{77.53}  \\
    \bottomrule
    \end{tabular}
    \vspace{-0.1in}
    \caption{\small Comparison of projector modules in inference latency and benchmark performance. Symbols $\downarrow$/$\uparrow$ indicate lower/higher is preferable. The Hybrid Compressor achieves superior accuracy with moderate latency overhead.}
    \label{tab:projector_efficiency}
     \vspace{-0.15in}
\end{table}

\begin{figure}[t]
  \centering
  \includegraphics[width=0.8\linewidth]{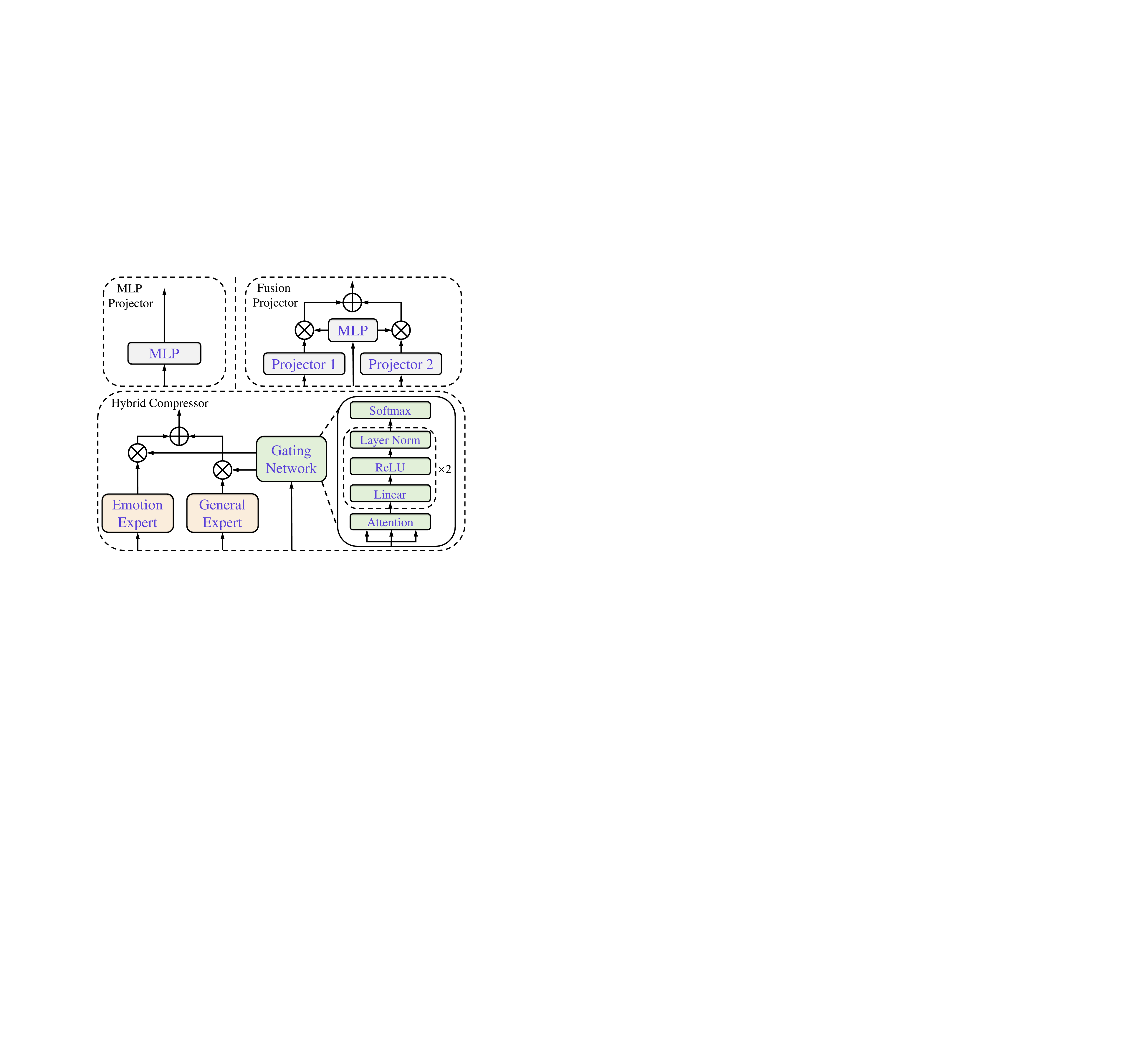}
\vspace{-0.1in}
\caption{\small Illustration of different projector architectures evaluated in Emotion-Qwen: MLP Projector, Fusion Projector, and the proposed Hybrid Compressor with a Gating Network.}
  \label{fig:projector}
\end{figure}

\noindent \textbf{Hybrid Compressor.} To evaluate the effectiveness of our proposed Hybrid Compressor, we conduct an ablation study comparing three projector configurations: the MLP projector, Fusion projector, and our Hybrid Compressor, as illustrated in Figure~\ref{fig:projector}. Specifically, the MLP projector employs two linear layers with a ReLU activation, while the Fusion projector comprises two parallel projectors whose outputs are dynamically combined via an MLP-based fusion ratio.

We follow the pre-training pipeline and utilize datasets listed in Table~\ref{tab:Pre-training data}. Results presented in Table~\ref{tab:Ablation on FEC Module and Hybrid Compressor} demonstrate that the Hybrid Compressor consistently outperforms alternative configurations. In particular, it achieves superior scores on MMBench, EmoSet, and MER2024 benchmarks, notably surpassing the MLP projector by over 8.44\% accuracy on the TextVQA benchmark.
Additionally, we evaluate the computational overhead and performance trade-offs associated with each projector setting, as detailed in Table~\ref{tab:projector_efficiency}. Despite the slight increase in parameter count and computational cost, our Hybrid Compressor achieves the optimal balance between accuracy and efficiency.
We further analyze the routing patterns of the Hybrid Compressor quantitatively by recording the average gating weights, as summarized in Table~\ref{tab:gating_weights}. The analysis indicates that the model dynamically assigns higher weights to the General Expert for general VL tasks and shifts towards the Emotion Expert for emotion-specific tasks, validating the adaptive and task-aware design of our Hybrid Compressor.

\begin{table}[t]
    \centering
    \setlength{\tabcolsep}{4.0pt}
    \begin{tabular}{lccc|cccc}
    \toprule
      \multicolumn{4}{c|}{General Benchmarks} & \multicolumn{4}{c}{Emotion Tasks} \\
    \cmidrule{1-8}
     \multicolumn{2}{c|}{MME}  & \multicolumn{2}{c|}{POPE}  & \multicolumn{2}{c|}{MER2024} & \multicolumn{2}{c}{EmoSet} \\
      \cmidrule(lr){1-2} \cmidrule(lr){3-4} \cmidrule(lr){5-6} \cmidrule(lr){7-8} 
      Emo & \multicolumn{1}{c|}{Gen}  & Emo & \multicolumn{1}{c|}{Gen} & Emo & \multicolumn{1}{c|}{Gen} & Emo & \multicolumn{1}{c}{Gen}  \\
    \midrule
     0.38 & \textbf{0.62} & 0.35 & \textbf{0.65}  & \textbf{0.63} & 0.37 & \textbf{0.57} & 0.43  \\
    \bottomrule
    \end{tabular}
    \vspace{-0.1in}
    \caption{\small Quantitative analysis of Hybrid Compressor’s routing patterns assigned to Emoiton Expert (Emo) and General Expert (Gen) across different task. We record the average gating weights assigned to two experts for each benchmark.}
    \label{tab:gating_weights}
    \vspace{-0.1in}
\end{table}

\begin{table}[t]
    \small
    \centering
    \setlength{\tabcolsep}{1.0pt}
    \begin{tabular}{l| cc cc cc c}
        \toprule
     \multirow{2}{*}{Strategy}  & \multicolumn{2}{c}{MER2024} & \multicolumn{2}{c}{DFEW}  & \multicolumn{2}{c}{EMER} & \multirow{2}{*}{EmoSet}\\
         &SEMI & NOISE & WAR & UAR& CLUE &LABEL\\
        \midrule
        FFT&82.77 & 76.60 &77.57& 57.36& 6.71 & 5.58  & 81.57 \\
        LoRA &  83.24 & 77.35 & 77.25 & 58.43 & 7.39 & 7.84 & 82.27 \\
        LoRAs& \textbf{85.47} & \textbf{79.67} & \textbf{78.31} & \textbf{62.11} & \textbf{8.25} & \textbf{8.16} & \textbf{85.49} \\
        \bottomrule
    \end{tabular}
    \vspace{-0.1in}
    \caption{\small Ablation Study of different training strategies. "FFT" denotes Full fine-tuning. "LoRA" refers to using single LoRA adapter, while "LoRAs" denotes the use of multiple adapters.}
    \label{tab:Ablation on Training Strategy}
    \vspace{-0.1in}
\end{table}

\begin{table}[t]
\small
    \centering
    \setlength{\tabcolsep}{4.0pt}
    \begin{tabular}{l|cc}
    \toprule
    Strategy  & Time Cost (hrs)$\downarrow$ & VRAM (GB)$\downarrow$ \\
    \midrule
    FFT&57.5 & 237.6\\
     Single LoRA & \textbf{17.0} &\textbf{153.9}\\
     Multi-LoRAs&17.8 &\textbf{153.9}\\
     \bottomrule
    \end{tabular}
    \vspace{-0.1in}
    \caption{\small Comparison of different training strategies on training time in hours (hrs) and total GPU VRAM usage in gigabytes (GB).}
    \label{tab:Comparison on Resource cost}
    \vspace{-0.1in}
\end{table}

\noindent \textbf{Training Strategy.} We systematically investigate various instruction-tuning strategies to optimize Emotion-Qwen, evaluating both task-specific performance and computational resource efficiency. Specifically, we compare three methodologies: Full fine-tuning, a single LoRA adapter, and multiple LoRA adapters customized individually for each emotion-specific dataset. Instruction-tuning datasets utilized in these experiments are detailed in the \textit{Emotional Instruction Fine-tuning} subsection. Results presented in Tables~\ref{tab:lora-finetuning-results} and \ref{tab:Comparison on Resource cost} demonstrate that employing multiple LoRA adapters achieves superior performance across all evaluated tasks, while maintaining comparable training times and identical GPU memory consumption relative to a single LoRA adapter. Conversely, full fine-tuning yields relatively inferior performance, likely due to the limited scale and distributional mismatch of our instruction-tuning datasets with the pre-trained LLM, exacerbating catastrophic forgetting effects. Overall, the use of multiple LoRA adapters provides the optimal balance between significantly improved task-specific accuracy and favorable computational efficiency.

\noindent \textbf{LLM Backbone.}
To evaluate the generalization capability of our proposed framework, we implemented Emotion-Qwen using two different LLM backbones. As summarized in Table~\ref{tab:different backbone}, our approach effectively integrates with LLaMA 3~\cite{Llama3}, achieving competitive results across all evaluated benchmarks.
Additionally, we observe that overall performance remains influenced by the underlying backbone’s capability. We plan to explore larger backbone models (e.g., 13B variants) in future work to further enhance performance.

\begin{table}[h]
\small
    \centering
    \setlength{\tabcolsep}{3.0pt}
    \begin{tabular}{l|c|c cc cc}
    \toprule
    \multirow{2}{*}{LLM}  & \multirow{2}{*}{Size} & \multirow{2}{*}{MME} & \multirow{1}{*}{Science} & \multirow{1}{*}{MER2024} & \multirow{2}{*}{EmoSet}  \\
    & & &-QA & SEMI/NOISE &\\
    
    \midrule
    LLaMA 3 & 7B & 1772.9&79.5&74.51/70.33&68.52\\
     Qwen 2.5 & 7B & 2163.5&77.2&82.85/77.53&81.54\\
     \bottomrule
    \end{tabular}
    \vspace{-0.1in}
    \caption{\small Comparison of different LLM backbone.}
    \label{tab:different backbone}
    \vspace{-0.1in}
\end{table}

\section{Conclusion}

In this work, we propose Emotion-Qwen, a unified multimodal framework explicitly designed to address critical limitations in existing emotional Large Multimodal Models (LMMs). Emotion-Qwen integrates a Facial Emotion Capture (FEC) module for precise extraction of expressive facial cues and a Hybrid Compressor composed of dual expert modules, effectively balancing emotion-specific representation and general vision-language reasoning. To comprehensively evaluate the model, we introduce the Video Emotional Reasoning (VER) dataset, a large-scale bilingual resource tailored for fine-grained, context-aware emotional reasoning.
Experimental results demonstrate that Emotion-Qwen achieves state-of-the-art performance across various emotion-centric benchmarks, while also maintaining superior results on general vision-language tasks. In summary, Emotion-Qwen sets a new standard for multimodal emotion understanding. We make our model weights and code publicly available to facilitate reproducibility and advance future research. Future studies will explore incorporating audio modalities and enhancing cross-modal generalization.


\bibliography{aaai2026}

\newpage

\end{document}